\documentclass{article}

\usepackage[preprint]{neurips_data_2024}

\usepackage[utf8]{inputenc} 
\usepackage[T1]{fontenc}    
\usepackage{hyperref}       
\usepackage{url}            
\usepackage{booktabs}       
\usepackage{amsfonts}       
\usepackage{nicefrac}       
\usepackage{microtype}      
\usepackage{xcolor}         
\usepackage{pifont}
\usepackage{multicol}
\usepackage{multirow}
\usepackage{colortbl}
\usepackage{array}
\usepackage{graphicx}
\usepackage{makecell}
\usepackage{caption}
\usepackage{subcaption}
\usepackage{longtable}
\usepackage{booktabs}

\newcommand{\dr}{\rowcolor[gray]{.95}}

\title{SegBook: A Simple Baseline and Cookbook for Volumetric Medical Image Segmentation}

\author{%
Jin Ye\textsuperscript{1}\thanks{Equal contribution.}
\qquad Ying Chen\textsuperscript{1,2}\footnotemark[1]
\qquad Yanjun Li\textsuperscript{1,3}
\qquad Haoyu Wang\textsuperscript{1}
\\
\textbf{Zhongying Deng\textsuperscript{4}}
\qquad \textbf{Ziyan Huang\textsuperscript{1}}
\qquad \textbf{Yanzhou Su\textsuperscript{1}}
\\
\textbf{Chenglong Ma\textsuperscript{1}}
\qquad \textbf{Yuanfeng Ji\textsuperscript{5}}
\qquad \textbf{Junjun He\textsuperscript{1\thanks{Corresponding author.}}}\\
\\
\textsuperscript{1}Shanghai AI Laboratory \quad \hspace{0.5em}  
\textsuperscript{2}Xiamen University\\ 
\textsuperscript{3}East China Normal University \quad \hspace{0.5em} 
\textsuperscript{4}University of Cambridge \quad \hspace{0.5em} 
\textsuperscript{5}Stanford University\\
{\tt\small \{yejin, hejunjun\}@pjlab.org.cn}
}

\begin{document}

\maketitle

\vspace{-0.5cm} 

\begin{center}
    \raisebox{-0.2\height}{\includegraphics[width=0.04\textwidth]{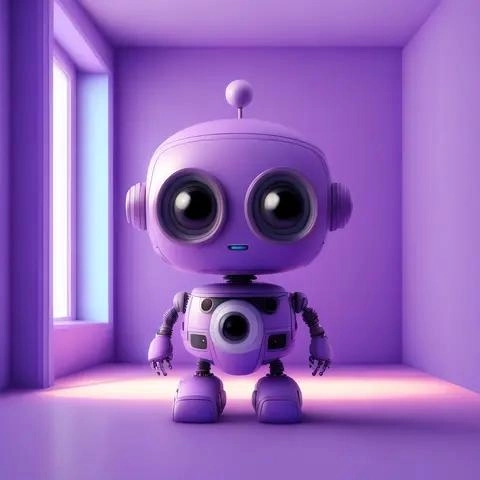}} 
    \hspace{0.2em} 
    \textbf{Project Page:} 
    \href{https://uni-medical.github.io/SegBook/index.html}{\texttt{https://uni-medical.github.io/SegBook/index.html}}
\end{center}

\vspace{1cm} 

\begin{abstract}
Computed Tomography (CT) is one of the most popular modalities for medical imaging. By far, CT images have contributed to the largest publicly available datasets for volumetric medical segmentation tasks, covering full-body anatomical structures. Large amounts of full-body CT images provide the opportunity to pre-train powerful models, e.g., STU-Net pre-trained in a supervised fashion, to segment numerous anatomical structures. However, it remains unclear in which conditions these pre-trained models can be transferred to various downstream medical segmentation tasks, particularly segmenting the other modalities and diverse targets. 
To address this problem, a large-scale benchmark for comprehensive evaluation is crucial for finding these conditions. Thus, we collected 87 public datasets varying in modality, target, and sample size to evaluate the transfer ability of full-body CT pre-trained models. We then employed a representative model, STU-Net with multiple model scales, to conduct transfer learning across modalities and targets. 
Our experimental results show that (1) there may be a bottleneck effect concerning the dataset size in fine-tuning, with more improvement on both small- and large-scale datasets than medium-size ones. (2) Models pre-trained on full-body CT  demonstrate effective modality transfer, adapting well to other modalities such as MRI. (3) Pre-training on the full-body CT not only supports strong performance in structure detection but also shows efficacy in lesion detection, showcasing adaptability across target tasks. We hope that this large-scale open evaluation of transfer learning can direct future research in volumetric medical image segmentation. 
\end{abstract}







\section{Introduction} 


Computed Tomography (CT) has emerged as one of the most indispensable modalities in medical imaging~\cite{panayides2020ai}, primarily due to its ability to provide detailed cross-sectional images of the human body.  Notably, CT images have contributed to the largest publicly available datasets for volumetric medical segmentation tasks, like TotalSegementor~\cite{totalsegmentator} which covers full-body anatomical structures. The comprehensive coverage of full-body anatomical structures as well as a large amount of image numbers can possibly provide full-body prior knowledge that is invaluable for a holistic understanding of human anatomy. Such prior knowledge is beneficial for transferring to a variety of downstream segmentation tasks. It thus offers us the opportunity to utilize full-body CT to explore models' transfer capabilities in various downstream medical segmentation tasks, particularly segmenting the other modalities and diverse targets. \emph{This problem is interesting because if these full-body CT-based models can be well transferred, full-body CT data can potentially serve as foundations for a wide range of medical segmentation tasks}. 

To investigate the conditions under which these pre-trained models can be transferred to various downstream medical segmentation tasks, we first identify key factors likely to impact transfer effectiveness, such as dataset size, modality, segmentation target, and model size. The first three factors (dataset size, modality, and segmentation target) require a benchmark with large-scale dataset size, various modalities, and diverse segmentation targets for comprehensive evaluations of these factors. The last factor, model sizes, necessitates a model with adjustable parameter scales. Scalable and Transferable U-Net (STU-Net)~\cite{stunet} is thus an ideal candidate due to its scalable model sizes and impressive transfer abilities to downstream tasks.


However, no large-scale evaluation has been conducted to benchmark the efficacy of transfer learning in volumetric medical image segmentation, leaving some important questions unanswered, such as the impact of dataset scales, the role of modality/target in transfer ability, and so on. Therefore, in this work, we first collected 87 public datasets varying in modality (\textit{e.g.}, CT, MRI, PET), target (\textit{e.g.}, structure, lesion), and size (small, medium, large). We then employed STU-Net of different scales to conduct a large-scale evaluation of transfer learning, focusing on modality and target transfer capabilities and the influence of dataset scales and model scales. We evaluate model transferability by training from scratch and fine-tuning pre-trained models on selected public datasets. Additionally, to demonstrate the level of performance in diverse tasks, we chose the significant training-from-scratch task-specific baseline nnU-Net~\cite{nnunet} for comparison.

Looking beyond STU-Net, our work benchmarks the transferability of large-scale supervised fully-body CT pre-training across 87 downstream datasets with diverse medical scenarios. We hope that the findings can provide insights for advancing transfer learning in volumetric medical images.


In summary, our key findings include:

\begin{itemize}
 
\item When fine-tuning the model on datasets of different sizes, interestingly, we observed the improvement in model performance does not increase linearly with the size of the dataset. Specifically, fine-tuning significantly enhanced the performance on both small and large datasets, whereas the improvement was relatively modest on medium-sized datasets. This indicates that there may be a bottleneck effect in which the benefits of fine-tuning decrease at certain data scales.
\item Models pre-trained on the CT with the full-body structure can effectively transfer to other modalities such as MRI, regardless of whether the downstream task targets appeared during the pre-training. This indicates the model's strong modality transfer capabilities.
\item Pre-training on CT structural data demonstrates promising performance in structure detection and efficacy in lesion detection, showcasing adaptability to various medical targets.
\end{itemize}

\section{Materials and Methods}

\subsection{Downstream Datasets}

\begin{figure}[htbp]
    \centering
    \begin{minipage}{0.37\textwidth}
        \centering
        \includegraphics[width=\textwidth]{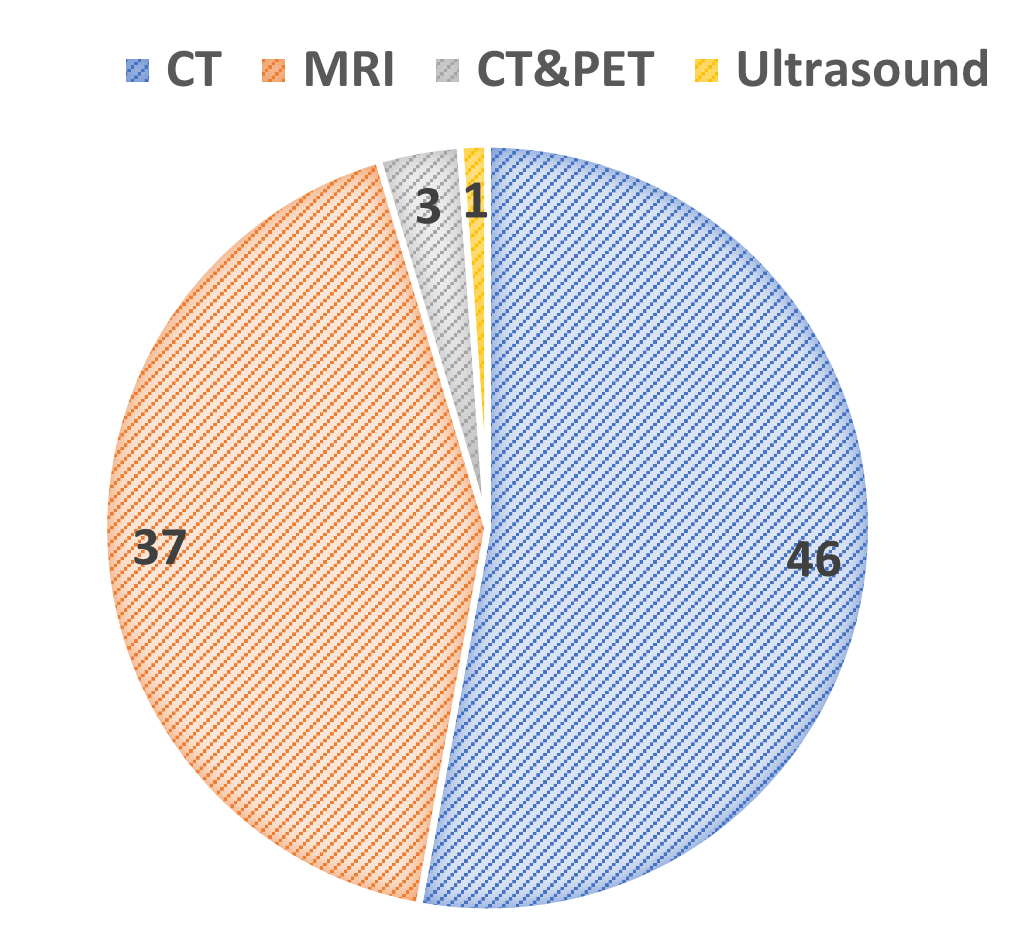}
        \caption{Numbers of datasets with different modalities.}
        \label{fig:fig1}
    \end{minipage}
    \hspace{0.05\textwidth} 
    \begin{minipage}{0.46\textwidth}
        \centering
        \includegraphics[width=\textwidth]{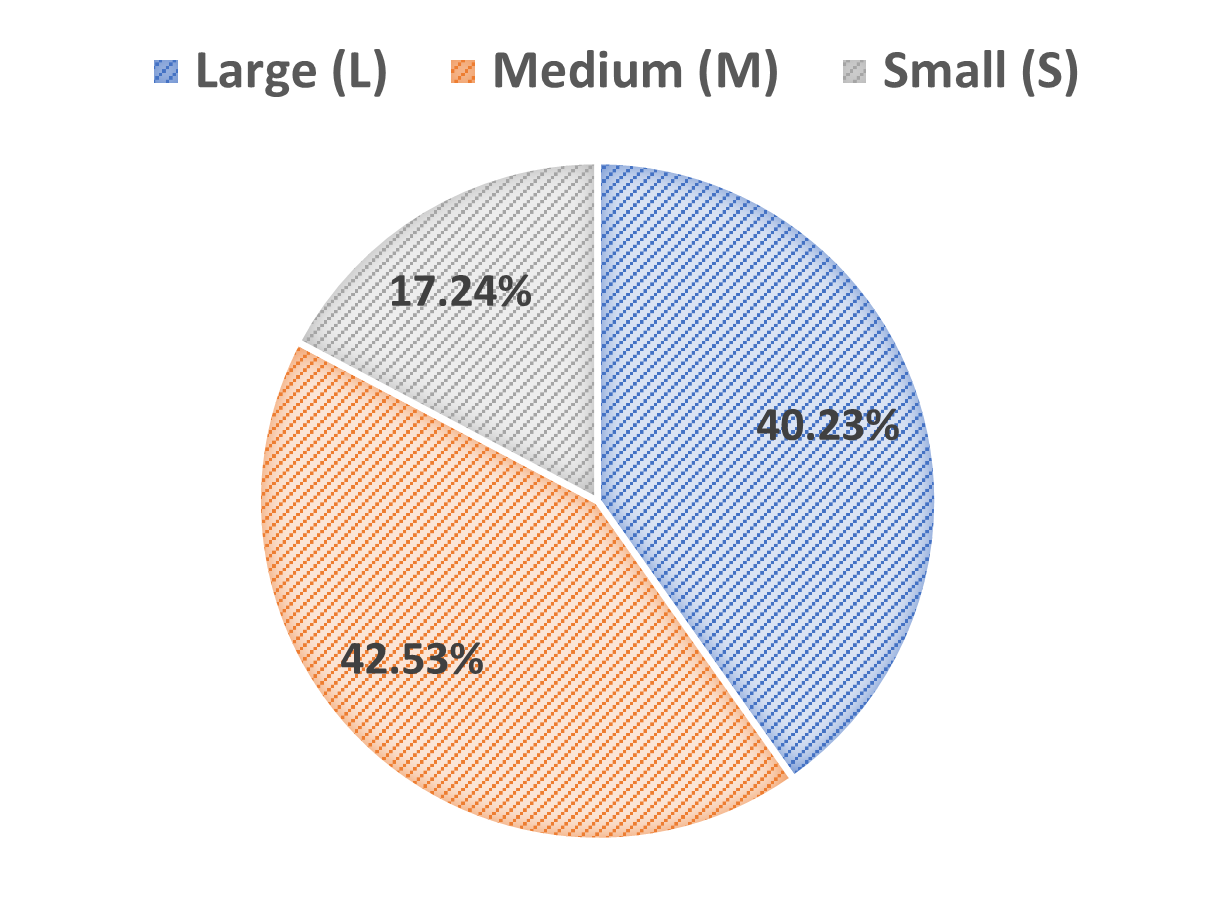}
        \caption{Proportions of datasets in different scales.}
        \label{fig:data_scale}
    \end{minipage}
\end{figure}

We assembled a comprehensive collection of 87 public datasets, which span diverse data sizes, targets, and modalities, to conduct an extensive evaluation, aimed at benchmarking the efficacy of transfer learning in volumetric medical image segmentation. These datasets were meticulously curated to support segmentation tasks targeting structures, lesions, and their combinations, encapsulating a wide range of imaging modalities, including CT, MRI, PET, and Ultrasound (US), as well as integrated modalities such as CT\&PET. Additionally, structures can be further categorized into seen structures and unseen structures: the former refers to those that appeared in the training set, while the latter refers to those that did not. Table~\ref{tab:datasets_num} provides a detailed summary of the distribution of datasets across various targets and modalities. Furthermore, these datasets exhibit a substantial variation in data scale, ranging from as few as 11 cases to as many as 1,250 cases per dataset, thereby offering a broad spectrum of data characteristics to robustly assess the performance of transfer learning strategies in medical imaging. We chose 100 as the boundary between small and medium categories and selected 400 as the boundary between medium and large categories. These categories respectively account for  40.23\%, 42.53\%, and 17.24\% of the total dataset population, as detailed in Figure~\ref{fig:data_scale}. This categorization facilitates a structured analysis of transfer learning performance across varying dataset complexities.

\begin{figure*}[h]
  \centering
  \includegraphics[width=0.85\linewidth]{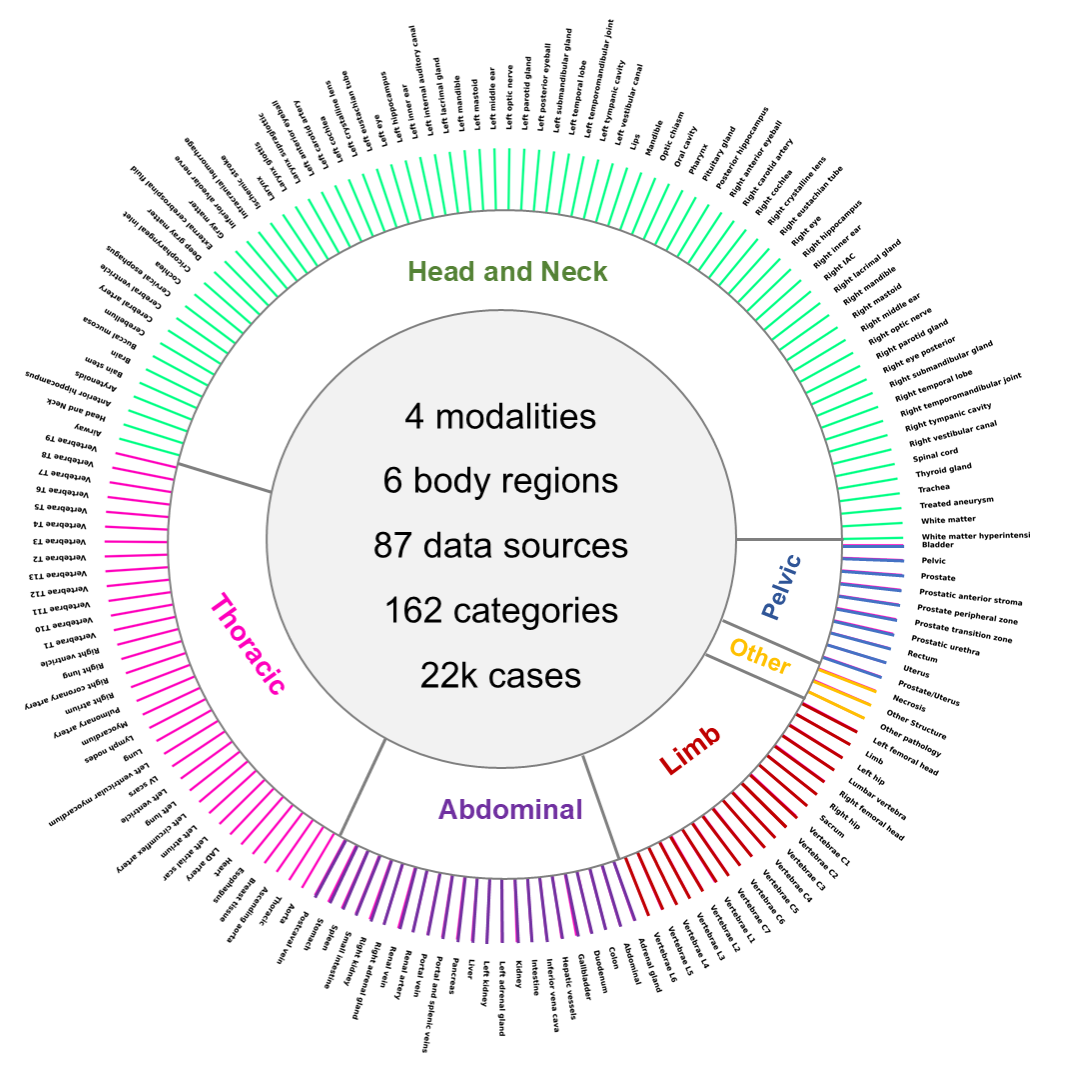}
  \caption{Overview of 87 datasets.}
  \label{fig:dataset}
\end{figure*}


\begin{table}[htbp]
\caption{Summary of number of datasets across targets and modalities.}
\label{tab:datasets_num}
\centering
\resizebox{\textwidth}{!}{
\begin{tabular}{c | c c c c | c c c c | c | c | c}
\toprule
Modality & \multicolumn{4}{c|}{CT} & \multicolumn{4}{c|}{MRI} & \multicolumn{1}{c|}{CT\&PET} & \multicolumn{1}{c|}{Ultrasound} &  \multirow{2}{*}{Total} \\
\cline{1-11}
Target & Lesion & \makecell[c]{Seen\\Sturctute} & \makecell[c]{Unseen\\Structure} & \makecell[c]{Lesion\\\&Sturcture} & Lesion & \makecell[c]{Seen\\Sturctute} & \makecell[c]{Unseen\\Structure} & \makecell[c]{Lesion\\\&Sturcture} & Lesion & Lesion \\\hline
datsets & 11 & 17 & 9 & 9 & 12 & 9 & 9 & 7 & 3 & 1 & 87\\
\bottomrule
\end{tabular}
}
\end{table}

\subsection{Pretrained Models}

To benchmark transfer learning with full-body CT supervised pretraining, we select STU-Net~\cite{stunet}, pre-trained on a large-scale full-body CT dataset, Totalsegmentator~\cite{totalsegmentator}. STU-Net introduces a flexible architecture that scales adeptly from lightweight to super-large configurations, catering to a broad spectrum of computational demands. The largest variant, STU-Net-H, boasts a staggering 1.4 billion parameters, demonstrating the model's exceptional scalability and robustness. STU-Net enhances the traditional convolutional block of nnU-Net by varying the network's depth and width, which results in four distinct configurations: small, base, large, and huge. 

The pre-trained weights of STU-Net we used are trained in a fully supervised manner for 4000 epochs on the TotalSegmentator dataset, which includes 1204 CT images annotated across 104 structures~\cite{wasserthal2023totalsegmentator}. This strategic utilization equips STU-Net with robust foundational knowledge, enabling it to excel in various downstream tasks with minimal fine-tuning.

In this work, we assessed the impact of pre-trained STU-Net across three different model scales: base, large, and huge. To highlight the performance variability across various tasks, we also included the well-established nnU-Net~\cite{nnunet} as a task-specific baseline, which was trained from scratch for comparison.

\subsection{Transfer Learning Setup}
We trained and evaluated STU-Net across 87 public datasets under the setting of supervised pre-training followed by fine-tuning.  
During fine-tuning on the downstream datasets, we initialize the segmentation head randomly while maintaining the pre-trained weights in the remaining layers, with all weights being learnable. The segmentation head uses 10$\times$ learning rate compared to other layers. To facilitate the comprehensive comparison, we also provide the results of training from scratch as baselines. Our experiments utilize the nnU-Net framework~\cite{nnunet}, with a patch-based training and sliding-window inference approach. Data pre-processing and augmentation for specific downstream tasks follow the automatic settings of nnU-Net. 
 
\subsection{Evaluation Metric}
We employ the Dice Similarity Coefficient (DSC) as our primary metric to evaluate segmentation performance. We formulate the definitions of the DSC as follows:
$$
DSC(G, P)=\frac{2|G \cap P|}{|G|+|P|},
$$
where $G$, $P$ denote the ground truth and the predicted mask, respectively. The DSC measures the overlap ratio between $P$ and $G$, where a higher DSC value signifies enhanced segmentation outcomes.

\section{Experiments and Analysis}
In this section, we present the outcomes of a comprehensive set of experiments designed to evaluate the effectiveness of transfer learning in volumetric medical image segmentation. We investigate the benefit of fine-tuning in diverse scenarios, exploring how model size, dataset scale,  modality, and target influence transfer capabilities.
 
\subsection{Overall Performance}

\begin{table}[!htb]
\caption{Dice Scores were calculated across 87 downstream datasets at different data scales: small (S), medium (M), and large (L). The symbol $\Delta (\cdot)$ denotes the improvement attributed to Pre-training (PT). }
\label{tab:data_scale_overall}
\centering
\begin{tabular}{l c c c| c c c | c }
\toprule
\multirow{2}{*}{\textbf{Method}} & \multirow{2}{*}{\textbf{PT}} & \multirow{2}{*}{\textbf{Params}} & \multirow{2}{*}{\textbf{TFLOPs}}& \multicolumn{3}{c|}{\textbf{Dataset Scale}} & \multirow{2}{*}{\textbf{Average}} \\

\cline{5-7} ~ & ~ & ~ & ~ & S & M & L \\ 
\midrule
nnU-Net      & ~         & $\sim$31M & $\sim$0.54 & 74.83	&	74.47	&	68.73	&	73.62
 \\ 
\midrule
STU-Net-base  & ~         & 58M      & 0.51  & 73.96	&	74.84	&	70.05	&	73.66
   \\ 
 STU-Net-base  & \ding{52} & 58M  & 0.51  & 76.17	&	76.59	&	72.81	&	75.77
 \\ 
\dr ~~$\Delta (base)$        & ~ & ~ & ~     & 2.21	&	1.75	&	2.76	&	2.11
   \\
STU-Net-large & ~         & 440M     & 3.81  & 74.14	&	75.71	&	70.48	&	74.18
   \\ 
 STU-Net-large & \ding{52} & 440M & 3.81  & 77.05	&	77.23	&	73.84	&	76.57
 \\ 
\dr ~~$\Delta (large)$       & ~ & ~ & ~     & 2.91	&	1.52	&	3.36	&	2.39
  \\
STU-Net-huge  & ~         & 1.4B     & 12.60 & 73.55	&	75.2	&	70.55	&	73.73
  \\ 
 STU-Net-huge  & \ding{52} & 1.4B & 12.60 & 76.87	&	77.14	&	74.21	&	76.53  \\ 
\dr ~~$\Delta (huge)$        & ~ & ~ & ~     & 3.32	&	1.94	&	3.66	&	2.80 \\
\bottomrule
\end{tabular}
\end{table}

\begin{figure*}[h]
  \centering
  \includegraphics[width=0.65\linewidth]{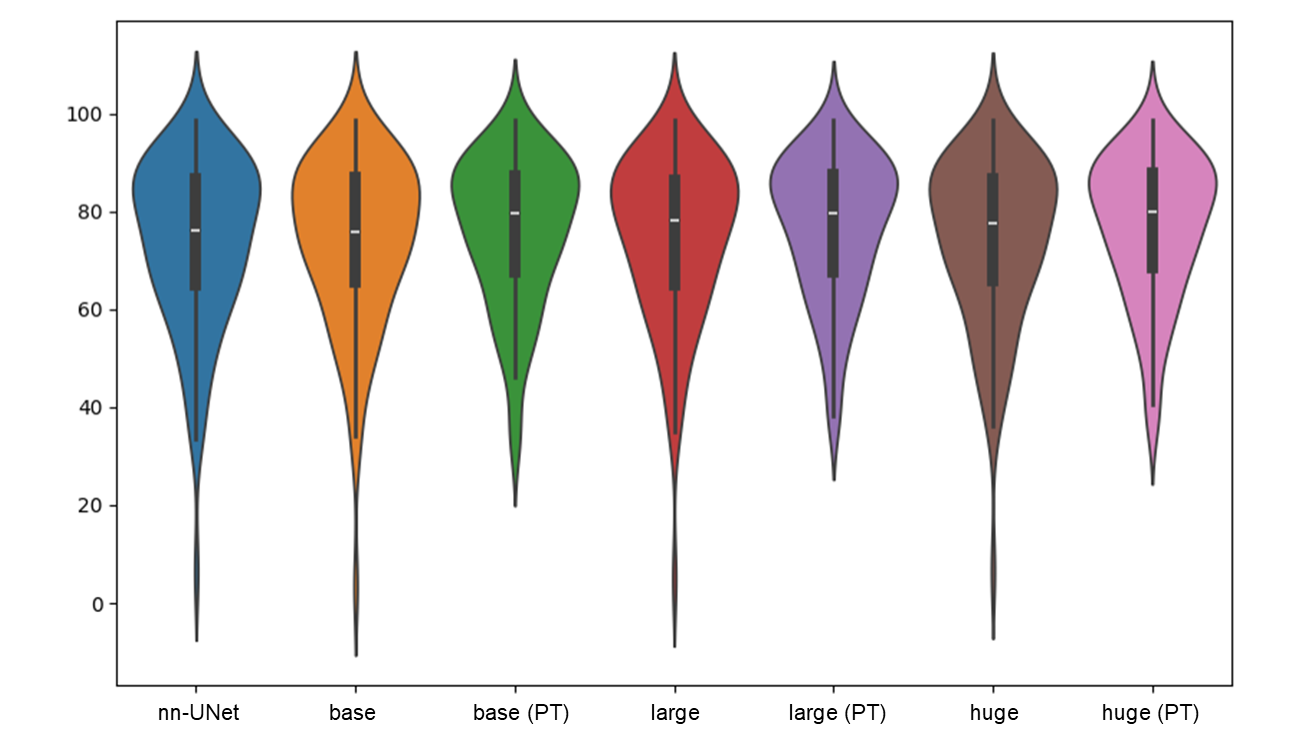}
  \caption{Violin plot for DSC for all 87 datasets with STU-Net in different scales.}
  \label{fig:Violin plot}
\end{figure*}

To investigate the overall transfer learning performance of STU-Net, we conducted experiments on 87 downstream datasets. As shown in Table~\ref{tab:data_scale_overall}, we reported average DSC for various models and detailed DSC for three data scales: small (S), medium (M), and large (L). The models evaluated include nnU-Net, STU-Net-base, STU-Net-large, and STU-Net-huge, with and without pre-training (PT). 

With pre-training, the performance of STU-Net at different sizes consistently increases under different dataset scale settings. Notably, the average performance of larger models (\textit{i.e.}, STU-Net-large and STU-Net-huge) as well as their performance across different data scales is better than smaller models (\textit{i.e.}, nnU-Net and STU-Net-base). In addition, we have observed an intriguing phenomenon where models of various scales exhibit significantly higher performance gains when fine-tuned on both small- and large-scale datasets, approximately around 3\%. However, on medium-scale datasets, the fine-tuning gains are only about 1\%. It appears that the benefits of fine-tuning do not proportionally increase with the scale of the dataset.

Figure~\ref{fig:Violin plot} presents violin plots comparing the DSC distributions across 87 datasets for different scales of the STU-Net model. The results highlight that models with pre-training (e.g., base(PT) and large(PT)) consistently achieve higher median DSC scores with less variance, demonstrating the effectiveness of fine-tuning. While larger models such as large and huge show competitive median performance, their broader distributions indicate some instability across datasets. Fine-tuning these models improves both stability and overall performance. 

\noindent \textbf{Summary: We observed a clear enhancement in the model's performance across various dataset sizes through fine-tuning techniques. Interestingly, the improvements were substantial for both small and large datasets, but are relatively less pronounced for medium-sized datasets. This suggests a non-linear relationship between dataset size and performance gains, implying a bottleneck effect where the benefits of fine-tuning diminish beyond a certain data scale.}

\subsection{Transfer across Targets}

\begin{table*}[htb]
\caption{Evaluation on the transferability across imaging modalities with STU-Net.}
\label{tab:result}
\centering
\resizebox{\textwidth}{!}{
\begin{tabular}{l|c|c c c|c c c | c |c}
\toprule
\multirow{3}{*}{\textbf{Method}} & \multirow{2}{*}{\textbf{PT}} & \multicolumn{3}{c}{\textbf{CT}} & \multicolumn{3}{c|}{\textbf{MRI}} & \textbf{US} & \textbf{CT\&PET} \\
\cline{3-10} ~ & ~ & \makecell[c]{Seen\\Structure} & \makecell[c]{Unseen\\Structure} & Lesion & \makecell[c]{Seen\\Structure} & \makecell[c]{Unseen\\Structure} & Lesion & Lesion & Lesion \\ 
\midrule
nnU-Net      & ~         & 82.28 & 69.59 & 58.86  &  87.16 & 83.27 & 68.50 & 49.66 & 58.88 \\
\midrule
STU-Net-base  & ~         & 82.79 & 70.06 & 58.88  &  86.67 & 82.20 & 68.15 & 53.70 & 62.44 \\
STU-Net-base  & \ding{52} & 85.00 & 73.85 & 63.14  &  87.47 & 82.62 & 69.14 & 54.54 & 66.45 \\
\dr ~~$\Delta (base)$ & ~       & 2.21 & 3.79 & 4.26 &  0.80 & 0.42  & 0.99 & 0.84 & 4.01 \\
STU-Net-large & ~         & 83.60 & 70.77 & 59.27 &  87.03 & 81.86 & 68.43 & 50.18 & 63.09 \\
 STU-Net-large & \ding{52} & 85.87 & 75.81 & 63.83 &  87.58 & 82.89 & 69.70 & 52.65 & 68.33 \\
\dr ~~$\Delta (large)$  & ~     & 2.27 & 5.04 & 4.56 &  0.55 & 1.03 & 1.27 & 2.47 & 5.24 \\
STU-Net-huge  & ~          & 82.73 & 70.67 & 57.35 &  86.94 & 82.11 & 68.54 & 49.38  & 62.83  \\
 STU-Net-huge  & \ding{52} & 85.90 & 75.15 & 63.61 &  87.59 & 83.49 & 69.45 & 52.78 & 68.74 \\
\dr ~~$\Delta (huge)$  & ~      & 3.17 & 4.48 & 6.26 & 0.65 & 1.38  & 0.91 & 3.40 & 5.91  \\
\bottomrule
\end{tabular}
}
\end{table*}


A common scenario arises when an evaluation dataset from the same modality (i.e., CT) is available, but the target of interest differs between the upstream and downstream tasks. In this case, we evaluate STU-Net's ability to effectively generalize across these varying targets, as shown in Table~\ref{tab:result}. The experimental results demonstrate that fine-tuning significantly enhances the performance of STU-Net models on both structure and lesion segmentation tasks.

Across the different model scales (base, large, huge), the training-from-scratch performance of STU-Net remains relatively consistent, indicating that model size alone does not drastically affect baseline performance. However, when it comes to fine-tuning improvements, larger models tend to show greater enhancements and higher gains.

\noindent \textbf{Summary: Pre-training on CT structure data not only yields excellent results in structure detection but also shows effective lesion detection, showcasing a notable degree of adaptability to the target task.}


\subsection{Transfer across Modalities}

\subsubsection{Modality Transfer with Seen Targets}
Table~\ref{tab:result} presents the performance of the STU-Net model across different imaging modalities. We first focus on the transfer from CT to MRI, where the structure has been seen during CT-based pre-training. 
The findings demonstrate STU-Net's ability to effectively generalize across modalities when the target category remains consistent. We also observed that, across models of various scales, fine-tuning consistently leads to an improvement in DSC compared to training from scratch, demonstrating the effectiveness of modal transfer with seen targets.

\subsubsection{Modality Transfer with Unseen Targets}

Table~\ref{tab:result} also demonstrates the transferability of models across different modalities and targets, specifically evaluating the DSC on datasets with unseen modalities and targets. We find that, despite targets being unseen, fine-tuning still leads to an improvement in DSC compared to training from scratch across all modalities and model scales, demonstrating the strong ability of modality transfer.

For single modalities, larger models also tend to show greater improvements from fine-tuning, which is more obvious to US transfer, with $\Delta (huge)$ of 0.84, 2.47, and 3.40 in STU-Net-base, STU-Net-large and STU-Net, respectively. It shows that larger models can be adapted to complex and diverse datasets more effectively.

\noindent \textbf{Summary: Models pre-trained on CT scans that include the full-body structure can effectively transfer to other modalities like MRI. This effectiveness holds regardless of whether the specific downstream task targets were present during pre-training, highlighting the model's strong modality transfer capabilities.}

\subsection{Transferability for Different Structures}

\begin{table*}[!htb]
\caption{Evaluation on the transferability across different structures.}
\label{tab:structure}
\centering
\resizebox{\textwidth}{!}{
\begin{tabular}{c | c | c c c c c c | c c }
\toprule
\textbf{Method} & \textbf{PT} & \textbf{Head and Neck} & \textbf{Pelvic}  & \textbf{Limb} & \textbf{Thoracic} & \textbf{Abdominal} & \textbf{other} & \textbf{Bone} & \textbf{Vessel} \\
\midrule
nnU-Net      & ~         & 79.16	&	84.2	&	62.9	&	73.15	&	89.52	&	63.84 & 65.47	&	80.46\\
\midrule
STU-Net-base  & ~         &  75.85	&	84.11	&	66.12	&	73.48	&	89.61	&	65.98 &  67.10	&	80.19\\
 STU-Net-base  & \ding{52} &  80.68	&	84.99	&	72.74	&	74.12	&	89.71	&	68.22 &  71.29	&	81.24\\
\dr ~~$\Delta (base)$        & ~ &  4.83	&	0.88	&	6.62	&	0.64	&	0.10	&	2.24 &  4.19	&	1.05 \\
STU-Net-large & ~         &  79.75	&	84.52	&	66.31	&	74.00	&	89.93	&	66.59 &  67.75	&	80.86\\
 STU-Net-large & \ding{52} &  81.51	&	85.15	&	74.16	&	74.80	&	90.28	&	68.82 &  72.39	&	82.06 \\
\dr ~~$\Delta (large)$       & ~ &  1.76	&	0.63	&	7.85	&	0.8	&	0.35	&	2.23 &  4.64	&	1.20 \\
STU-Net-huge  & ~         &  78.99	&	84.29	&	65.88	&	73.90	&	89.85	&	65.99 &  67.32	&	80.53 \\
 STU-Net-huge  & \ding{52} &  80.37	&	85.99	&	73.97	&	74.92	&	90.37	&	69.86 &  72.33	&	82.00 \\
\dr ~~$\Delta (huge)$        & ~ &  1.38	&	1.70	&	8.09	&	1.02	&	0.52	&	3.87 &  5.01	&	1.47 \\
\bottomrule
\end{tabular}
}
\end{table*}


As shown in Table~\ref{tab:structure}, models pre-trained on CT structural data demonstrate effective transferability across a wide range of anatomical structures. The consistent improvements across different regions, especially with fine-tuning, suggest that these pre-trained models capture meaningful representations that generalize well beyond their initial training data. STU-Net-large and STU-Net-huge models, in particular, show strong adaptability, achieving high performance in complex regions like the thoracic and abdominal structures. Although in certain regions, such as the pelvic area, the models exhibit limited gains ($\Delta$), the overall results validate the versatility of the pre-trained models with higher scores. This highlights the potential of using CT-based pre-training as a foundational step to enhance model generalization across diverse anatomical contexts, enabling broader applicability in medical imaging tasks.

\subsection{Transferability for Bone and Vessel}

Table~\ref{tab:structure} presents the evaluation results for bone and vessel segmentation tasks. Pre-training on CT structural data proves to be highly effective for transferring knowledge to complex segmentation tasks, such as bone and vessel segmentation. The consistent performance improvements across all models, especially after fine-tuning, demonstrate the robustness and adaptability of these pre-trained models. STU-Net-huge’s notable 5.01\% DSC gain in bone segmentation highlights that even challenging anatomical structures benefit significantly from pre-trained knowledge. This suggests that CT-based pre-training equips models with generalizable features, enabling efficient adaptation to various medical imaging tasks, including those requiring high precision in fine-grained regions like bones and vessels.







\section{Conclusion and Future Work}

\subsection{Conclusion}
In this study, we examine the transfer ability of full-body CT pre-trained models. We collected large-scale public datasets varying in modality, target, and size. We utilized STU-Net of different scales to conduct a series of transfer learning experiments for volumetric medical image segmentation. We compare overall performance to measure the effectiveness of modality and target transfers, leveraging various datasets and model scales. We hope that this large-scale evaluation of transfer learning can direct future research on volumetric medical image segmentation.

\subsection{Future Work}

In our research, we mainly explored full-body CT-structure pre-training transferring to other modalities and targets and observed strong effectiveness. In future work, we will consider exploring the transfer effect between more models and targets and also consider using different fine-tuning techniques for further exploration. We hope that our research can bring more guidance for pre-training and transfer learning in volumetric medical image segmentation.

\newpage
\begin{center}
\setlength\LTleft{0pt}
\setlength\LTright{0pt}
\begin{longtable}{lccc}
	\caption{Detailed datsets}
	\label{tab:dataset num} \\
	\toprule
    Dataset & Modality & Target & Case \\ \hline
Task001-BrainTumour	~\cite{antonelli2022medical}	&	MRI	&	Lesion	&	484	\\
Task002-Heart	~\cite{antonelli2022medical}	&	MRI	&	Seen Structure	&	20	\\
Task003-Liver	~\cite{antonelli2022medical}	&	CT	&	Structure\&Lesion	&	130	\\
Task004-Hippocampus	~\cite{antonelli2022medical}	&	MRI	&	Unseen Structure	&	260	\\
Task005-Prostate	~\cite{antonelli2022medical}	&	MRI	&	Seen Structure	&	31	\\
Task006-Lung	~\cite{antonelli2022medical}	&	CT	&	Lesion	&	63	\\
Task007-Pancreas	~\cite{antonelli2022medical}	&	CT	&	Structure\&Lesion	&	280	\\
Task008-HepaticVessel	~\cite{antonelli2022medical}	&	CT	&	Structure\&Lesion	&	303	\\
Task009-Spleen	~\cite{antonelli2022medical}	&	CT	&	Seen Structure	&	40	\\
Task010-Colon	~\cite{antonelli2022medical}	&	CT	&	Lesion	&	125	\\
Task011-BTCV	~\cite{landman2015miccai}	&	CT	&	Seen Structure	&	30	\\
Task012-BTCV-Cervix	~\cite{landman2015miccai}	&	CT	&	Seen Structure	&	30	\\
Task013-ACDC	~\cite{bernard2018deep}	&	MRI	&	Seen Structure	&	200	\\
Task019-BraTS21	~\cite{baid2021rsna}	&	MRI	&	Lesion	&	1250	\\
Task020-AbdomenCT1K	~\cite{ma2021abdomenct}	&	CT	&	Seen Structure	&	1000	\\
Task021-KiTS2021	~\cite{heller2023kits21}	&	CT	&	Structure\&Lesion	&	300	\\
Task023-FLARE22	~\cite{ma2023unleashing}	&	CT	&	Seen Structure	&	70	\\
Task029-LITS	~\cite{bilic2023liver}	&	CT	&	Structure\&Lesion	&	130	\\
Task034-Instance22	~\cite{li2023state}	&	CT	&	Unseen Structure	&	100	\\
Task036-KiPA22	~\cite{he2020dense}	&	CT	&	Structure\&Lesion	&	70	\\
\makecell[l]{Task037-CHAOS-Task-3-5-Variant1\\~\cite{kavur2021chaos}}		&	MRI	&	Seen Structure	&	40	\\
Task039-Parse22	~\cite{luo2023efficient}	&	CT	&	Seen Structure	&	100	\\
Task040-ATM22	~\cite{zhang2023multi}	&	CT	&	Unseen Structure	&	300	\\
\makecell[l]{Task041-ISLES2022\\	~\cite{hernandez2022isles}}	&	MRI	&	Lesion	&	250	\\
Task044-CrossMoDA23	~\cite{DORENT2023102628}	&	MRI	&	Structure\&Lesion	&	226	\\
Task044-KiTS23	~\cite{heller2023kits19}	&	CT	&	Structure\&Lesion	&	489	\\
Task050-LAScarQS22-task1	~\cite{li2022atrialjsqnet}	&	MRI	&	Seen Structure	&	60	\\
Task051-AMOS-CT	~\cite{ji2022amos}	&	CT	&	Seen Structure	&	300	\\
Task051-LAScarQS22-task2	~\cite{li2022atrialjsqnet}	&	MRI	&	Seen Structure	&	130	\\
Task052-AMOS-MR	~\cite{ji2022amos}	&	MRI	&	Seen Structure	&	60	\\
Task053-AMOS-Task2	~\cite{ji2022amos}	&	MRI	&	Seen Structure	&	360	\\
Task083-VerSe2020	~\cite{sekuboyina2021verse}	&	CT	&	Seen Structure	&	350	\\
Task103-ADAM2020	~\cite{fang2022adam}	&	MRI	&	Structure\&Lesion	&	113	\\
\makecell[l]{Task104-Colorectal-Liver-Metastases\\	~\cite{simpson2024preoperative}}	&	CT	&	Structure\&Lesion	&	196	\\
\makecell[l]{Task105-DICOM-LIDC-IDRI-Nodules\\	~\cite{fedorov2018standardized}}	&	CT	&	Unseen Structure	&	1018	\\
Task106-AIIB2023	~\cite{nan2023fuzzy}	&	CT	&	Unseen Structure	&	120	\\
Task107-HCC-TACE-Seg	~\cite{moawad2021multimodality}	&	CT	&	Structure\&Lesion	&	224	\\
Task108-ISBI-MR-Prostate-2013	~\cite{bloch2015}	&	MRI	&	Unseen Structure	&	79	\\
Task109-SMILE-UHURA2023	~\cite{smileuhura2023}	&	MRI	&	Unseen Structure	&	11	\\
\makecell[l]{Task110-ISPY1-Tumor-SEG-Radiomics\\~\cite{chitalia2022expert}}	&	MRI	&	Lesion	&	160	\\
\makecell[l]{Task111-LUAD-CT-Survival\\	~\cite{goldgof2017long}}	&	CT	&	Lesion	&	40	\\
\makecell[l]{Task112-PROSTATEx-Seg-HiRes\\	~\cite{schindele2020high}}	&	MRI	&	Unseen Structure	&	65	\\
\makecell[l]{Task113-PROSTATEx-Seg-Zones\\	~\cite{schindele2020high}}	&	MRI	&	Unseen Structure	&	98	\\
\makecell[l]{Task114-Prostate-Anatomical-Edge-Cases\\~\cite{thompson2023}}		&	CT	&	Seen Structure	&	130	\\
Task115-QIBA-VolCT-1B	~\cite{mcnitt2015determining}	&	CT	&	Lesion	&	149	\\
Task116-ISPY1	~\cite{chitalia2022expert}	&	MRI	&	Structure\&Lesion	&	820	\\
\makecell[l]{Task166-Longitudinal Multiple Sclerosis\\ Lesion Segmentation~\cite{carass2017longitudinal}}	&	MRI	&	Lesion	&	20	\\
Task502-WMH	~\cite{kuijf2019standardized}	&	MRI	&	Unseen Structure	&	60	\\
Task503-BraTs2015	~\cite{BRATS15}	&	MRI	&	Structure\&Lesion	&	274	\\
Task504-ATLAS	~\cite{labella2023asnrmiccai}	&	MRI	&	Lesion	&	655	\\
Task507-Myops2020	~\cite{luo2022mathcal}	&	MRI	&	Structure\&Lesion	&	25	\\
Task511-ATLAS2023	~\cite{quinton2023tumour}	&	MRI	&	Structure\&Lesion	&	60	\\
Task525-CMRxMotions	~\cite{wang2022extreme}	&	MRI	&	Seen Structure	&	139	\\
Task556-FeTA2022-all	~\cite{payette2021automatic}	&	MRI	&	Unseen Structure	&	120	\\
Task559-WORD	~\cite{luo2022word}	&	CT	&	Seen Structure	&	120	\\
Task601-CTSpine1K-Full	~\cite{deng2021ctspine1k}	&	CT	&	Seen Structure	&	1005	\\
Task603-MMWHS	~\cite{gonzalez2019}	&	CT	&	Seen Structure	&	40	\\
Task605-SegThor	~\cite{lambert2020segthor}	&	CT	&	Seen Structure	&	40	\\
Task606-orCaScore	~\cite{wolterink2016evaluation}	&	CT	&	Unseen Structure	&	31	\\
Task611-PROMISE12	~\cite{litjens2014evaluation}	&	MRI	&	Unseen Structure	&	50	\\
Task612-CTPelvic1k	~\cite{liu2021deep}	&	CT	&	Seen Structure	&	1105	\\
Task613-COVID-19-20	~\cite{roth2022rapid}	&	CT	&	Lesion	&	199	\\
Task614-LUNA16	~\cite{setio2017validation}	&	CT	&	Unseen Structure	&	888	\\
Task615-Chest-CT-Scans-with-COVID-19	~\cite{}	&	CT	&	Lesion	&	50	\\
Task616-LNDb	~\cite{pedrosa2019lndb}	&	CT	&	Lesion	&	235	\\
\makecell[l]{Task628-StructSeg2019-subtask1\\	~\cite{heimann2009comparison}}	&	CT	&	Unseen Structure	&	50	\\
\makecell[l]{Task629-StructSeg2019-subtask2\\	~\cite{heimann2009comparison}}	&	CT	&	Seen Structure	&	50	\\
\makecell[l]{Task630-StructSeg2019-subtask3\\	~\cite{heimann2009comparison}}	&	CT	&	Lesion	&	50	\\
\makecell[l]{Task631-StructSeg2019-subtask4\\	~\cite{heimann2009comparison}}	&	CT	&	Lesion	&	50	\\
Task666-MESSEG	~\cite{styner20083d}	&	MRI	&	Lesion	&	40	\\
Task700-SEG-A-2023	~\cite{radl2022avt}	&	CT	&	Seen Structure	&	55	\\
Task701-LNQ2023	~\cite{lnq2023}	&	CT	&	Lesion	&	393	\\
Task701-SegRap2023	~\cite{luo2023segrap2023}	&	CT	&	Seen Structure	&	120	\\
Task702-CAS2023	~\cite{cas2023}	&	MRI	&	Unseen Structure	&	100	\\
Task702-SegRap2023-Task2	~\cite{luo2023segrap2023}	&	CT	&	Lesion	&	120	\\
Task703-TDSC-ABUS2023	~\cite{zhou2021multi}	&	Ultrasound	&	Lesion	&	100	\\
Task704-ToothFairy2023	~\cite{2022CVPR}	&	CT	&	Unseen Structure	&	153	\\
Task710-autoPET	~\cite{gatidis2022whole}	&	CT\&PET	&	Lesion	&	1014	\\
Task711-autoPET-PET-only	~\cite{gatidis2022whole}	&	CT\&PET	&	Lesion	&	500	\\
Task712-autoPET-CT-only	~\cite{gatidis2022whole}	&	CT\&PET	&	Lesion	&	1014	\\
Task720-HIE2023	~\cite{bao2023boston}	&	MRI	&	Lesion	&	85	\\
Task894-BraTS2023-MET	~\cite{moawad2023brain}	&	MRI	&	Lesion	&	238	\\
Task895-BraTS2023-SSA	~\cite{adewole2023brain}	&	MRI	&	Lesion	&	43	\\
Task896-BraTS2023-PED	~\cite{kazerooni2023brain}	&	MRI	&	Lesion	&	99	\\
Task898-BraTS2023-MEN	~\cite{labella2023asnrmiccai}	&	MRI	&	Lesion	&	1000	\\
Task899-BraTS2023-GLI	~\cite{menze2014multimodal}	&	MRI	&	Structure\&Lesion	&	1250	\\
Task966-HaN-Seg	~\cite{podobnik2023han}	&	CT	&	Unseen Structure	&	41	\\
\bottomrule
\end{longtable}
\end{center}

\bibliographystyle{abbrvnat}
\bibliography{main}

\end{document}